\documentclass[showkeys,superscriptaddress,floatfix,prd,10pt,aps]{revtex4-2}
\usepackage{graphicx,epstopdf}
\usepackage[caption=false]{subfig}
\usepackage{appendix}
\usepackage[T1]{fontenc}
\usepackage{lmodern}
\usepackage[dvipsnames,x11names]{xcolor}
\usepackage[colorlinks=true,linkcolor=ForestGreen,citecolor=ForestGreen,urlcolor=NavyBlue]{hyperref}
\usepackage[sort&compress]{natbib}
\usepackage{lipsum}
\usepackage{morefloats}
\usepackage[pdf]{pstricks}
\usepackage{amsmath}
\usepackage{amssymb}
\usepackage{amsfonts}
\usepackage{rotating}
\usepackage{cancel}
\usepackage{mathtools}
\usepackage{bbm}
\usepackage{dsfont}
\usepackage{bbold}
\usepackage{multirow}
\usepackage{ulem}
\usepackage{physics}
\usepackage{orcidlink}
\usepackage{colortbl}

\def\pslash{p\!\!\!\slash }
\def\qslash{q\!\!\!\slash }
\def\xslash{x\!\!\!\slash }
\def\eslash{\varepsilon\!\!\!\slash }

\def\vel{\left|}
\def\ver{\right|}

\begin{document}

\title{Magnetic dipole moments of bottom-charm baryons in light-cone QCD }

\author{Ula\c{s} \"{O}zdem\orcidlink{0000-0002-1907-2894}}%
\email[]{ulasozdem@aydin.edu.tr}
\affiliation{Health Services Vocational School of Higher Education, Istanbul Aydin University, Sefakoy-Kucukcekmece, 34295 Istanbul, T\"{u}rkiye}

\date{\today}  

\begin{abstract}
The magnetic dipole moments of the doubly-heavy baryons include significant data on their inner structure and geometric shape. Moreover, understanding the electromagnetic properties of doubly-heavy baryons is the key to confinement and heavy flavor effects.   Inspired by this, we extract the magnetic dipole moments of the spin-$\frac{1}{2}$  bottom-charm baryons utilizing the QCD light-cone sum rule with considering the distribution amplitudes of the photon. 
The magnetic dipole moments are obtained as $\mu_{\Xi_{bc}^{+}} = -0.50^{+0.14}_{-0.12}~\mu_{N}$, $\mu_{\Xi_{bc}^{0}} = 0.39^{+0.06}_{-0.05}~\mu_{N}$ and $\mu_{\Omega_{bc}^{0}} = 0.38^{+0.05}_{-0.04} ~\mu_{N}$, $\mu_{\Xi_{bc}^{\prime +}} = 0.57^{+0.13}_{-0.12}~\mu_{N}$, $\mu_{\Xi_{bc}^{\prime 0}} =-0.29^{+0.07}_{-0.06}~\mu_{N}$ and $\mu_{\Omega_{bc}^{\prime 0}} = -0.26^{+0.06}_{-0.05}~\mu_{N}$. Comparing the results obtained on the magnetic dipole moments of the $\Omega^{(\prime)0}_{bc}$ baryon with those of the $\Xi^{(\prime)0}_{bc}$ baryon, the $U$-symmetry is minimally broken.  We have compared our results with other theoretical predictions that could be a useful complementary tool for the interpretation of the doubly-heavy baryon sector, and we observe that they are not in mutual agreement with each other.
\end{abstract}

\maketitle

\section{Motivation}
Heavy quark baryons play an important role in hadron physics and the investigation of their features may deepen our comprehension of QCD. In the last few decades, measurements of heavy baryons have accelerated with improvements in experimental facilities. Numerous states have been discovered, but many states need confirmation; hence the heavy baryon sector is theoretically attractive.  One of the doubly-charmed baryons ($\Xi_{cc}^+$), which contains two c and one d quark,  was first announced by the SELEX Collaboration with the mass $M_{\Xi^+_{cc}}=3519 \pm 1~$MeV~\cite{Mattson:2002vu},  although, the result of SELEX could neither be confirmed nor excluded by other experimental groups such as FOCUS~\cite{Ratti:2003ez}, BABAR~\cite{Aubert:2006qw}, and Belle~\cite{Chistov:2006zj}. In 2017,  a doubly-charmed baryon $\Xi_{cc}^{++}$ with the mass $M_{\Xi_{cc}^{++}}=3621.40\pm 0.72\pm0.27\pm0.14~$MeV was discovered by the LHCb Collaboration~\cite{Aaij:2017ueg}.  The observation of the $\Xi_{cc}^{++}$ state opened the door to the experimental detection of doubly-heavy baryons, and scientists expected to discover more doubly-heavy baryons experimentally.  The LHCb collaboration has also carried out searches for the doubly heavy baryons, $\Xi^+_{bc}$,  $\Xi^0_{bc}$ and $\Omega_{bc}^0$, yet these baryons are yet to be observed \cite{LHCb:2020iko,LHCb:2021xba,LHCb:2022fbu}. Further measurements will be possible with larger data samples, and additional decay modes are expected at the upgraded LHCb experiments.
The study of doubly-heavy baryons contributes to an in-depth comprehension of the chiral dynamics, heavy quark symmetry,  fundamental theory of the strong interaction, and models inspired by QCD.  Therefore, numerous theoretical studies inspired by these discoveries also have been carried out to define the characteristic features like the masses, decay widths, strong coupling constants, and lifetimes of doubly-heavy baryons by investigating the strong and weak decays of these baryons within different theoretical models and approaches. Reviews of doubly-heavy baryon physics together with a more complete list of references can be found in Refs. \cite{Roberts:2007ni,Crede:2013kia,Chen:2022asf,Meng:2022ozq}. Electromagnetic properties of doubly-heavy baryons
have been studied in Refs. \cite{Ozdem:2018uue,Ozdem:2019zis,Bahtiyar:2022nqw,Mutuk:2021epz,Silvestre-Brac:1996myf,Albertus:2006ya,Gadaria:2016omw,Dhir:2013nka,Simonis:2018rld,Bernotas:2012nz,Shah:2016vmd,Shah:2017liu,Patel:2007gx,Li:2017cfz,Faessler:2006ft,Can:2013zpa,Can:2013tna,Patel:2008xs,Julia-Diaz:2004yqv,Sharma:2010vv,Zhang:2021yul,HillerBlin:2018gjw,Liu:2018euh,Li:2020uok,Meng:2017dni}.   It is worth noting that the different approaches used to derive the properties of double-heavy baryons lead to quite different estimations,  which can be used to distinguish between approaches.

Investigating the electromagnetic features of hadrons ensures us with helpful knowledge about their inner organizations. One can extract information about their shapes, sizes, and decay widths and compare them with the experimental results. Baryons including two heavy quarks are particularly stimulating to study since examining the electromagnetic features of two heavy quarks bound to a light quark helps us figure out the internal interaction dynamics of baryons containing heavy quarks. Moreover, the results can help us understand the key features of QCD such as confinement and flavor effects.  In the present study, the magnetic dipole moments of spin-$\frac{1}{2}$ bottom-charm baryons  ($B^{(\prime)}_{bc}$ for short) are obtained within the QCD light-cone sum rule (LCSR). This method is powerful for studying the dynamic and static properties of hadrons.  In the LCSR method~\cite{Chernyak:1990ag, Braun:1988qv, Balitsky:1989ry}, a relevant two-point correlation function within the external background electromagnetic field is evaluated in two different ways. Firstly, the so-called hadronic representation is obtained in terms of hadronic parameters such as form factors, magnetic dipole moments, etc. In the second way, the so-called QCD representation is evaluated in terms of quark-gluon degrees of freedom. The correlation functions acquired in these two different ways are then correlated using the assumption of quark-hadron duality. As a final step, Borel transform and continuum subtraction are applied to eliminate the contributions of possible higher states and continuum. In this way, the LCSR for the desired physical quantities, in our case magnetic dipole moments, are acquired.   

This work is organized as follows. In Sec. \ref{formalism}, the calculation method used in this work is briefly introduced.  We present our results, compare them to other works, and give a discussion in Sec. \ref{numerical}. Finally, a summary is given in Sec. \ref{summary}.

\begin{widetext}

\section{ The LCSR for magnetic dipole moments of the $B^{(\prime)}_{bc}$ baryons }\label{formalism}

As we mentioned above, the starting point to evaluate any physical quantity in the LCSR method is to write a convenient correlation function. Here for the magnetic dipole moments, the correlation function is written as
\begin{equation}
 \label{edmn01}
\Pi(p,q)=i\int d^{4}xe^{ip\cdot x}\langle 0|\mathcal{T}\{J(x)
\bar J(0)\}|0\rangle_{\gamma},
\end{equation}%
where q is the momentum of the photon, $\gamma$ is the external electromagnetic field
and $J(x)$ is the interpolating current having quantum numbers $J^P =\frac{1}{2}^+$.  
The interpolating current $J(x)$ is one of the key components necessary to extract the magnetic dipole moments of the bottom-charm baryons employing the LCSR method. There are different ways to construct the corresponding interpolating currents using the diquark-quark configurations of different spin parities. Taking into account all the properties of the  bottom-charm baryons, the following  interpolating currents are constructed in Ref. \cite{Wang:2022ufh}, that we employ to evaluate the magnetic dipole moments of these bottom-charm baryons:
\begin{align}
\label{cur1}
 J_{1}(x) &= \varepsilon^{abc} \big[b_a^T(x)C\gamma_\mu c_b(x)\big]\gamma_5 \gamma^\mu  q_c(x),\nonumber\\
  J_{2}(x) &= \varepsilon^{abc} \big[b_a^T(x)C\gamma_5 c_b(x)\big] q_c(x),
\end{align}
where q is u, d or s-quark, a, b, and c are color indices; and 
C is the charge conjugation matrix. However, detailed investigations show that the $J_1$ interpolating current is symmetric under the $b\leftrightarrow c$ exchange, while the $J_2$ interpolating current is antisymmetric under the $b\leftrightarrow c$ exchange.  Therefore, the $J_1$ interpolating current couples to the $\Xi_{bc}$ and $\Omega_{bc}$ baryons, while the $J_2$ interpolating current couples to the $\Xi_{bc}^{\prime}$ and $\Omega_{bc}^{\prime}$  baryons.

To evaluate the physical representation of the correlation function, we insert two complete sets of hadronic states having the same quantum numbers as that of bottom-charm baryons interpolating currents and we get:

\begin{eqnarray}\label{edmn02}
\Pi^{Had}(p,q)&=&\frac{\langle0\mid J (x)\mid
{B^{(\prime)}_{bc}}(p)\rangle}{[p^{2}-m_{{B^{(\prime)}_{bc}}}^{2}]}\langle {B^{(\prime)}_{bc}}(p)\mid
{B^{(\prime)}_{bc}}(p+q)\rangle_\gamma 
\frac{\langle {B^{(\prime)}_{bc}}(p+q)\mid
\bar{J} (0)\mid 0\rangle}{[(p+q)^{2}-m_{{B^{(\prime)}_{bc}}}^{2}]}+...,
\end{eqnarray}
where dots stand for the contribution of the continuum and higher states. As can be seen from in  Eq. (\ref{edmn02}), we need explicit forms of matrix elements such as $\langle0\mid J_{B^{(\prime)}_{bc}}(0)\mid {B^{(\prime)}_{bc}}(p,s)\rangle$ and $\langle {B^{(\prime)}_{bc}}(p)\mid {B^{(\prime)}_{bc}}(p+q)\rangle_\gamma$ and they are written as follows
\begin{eqnarray}\label{edmn03}
\langle0\mid J_{B^{(\prime)}_{bc}}(0)\mid {B^{(\prime)}_{bc}}(p,s)\rangle&=&\lambda_{{B^{(\prime)}_{bc}}}\nu(p,s),\\
\langle {B^{(\prime)}_{bc}}(p)\mid {B^{(\prime)}_{bc}}(p+q)\rangle_\gamma &=&\varepsilon^\mu \bar \nu(p)\Bigg[\big(F_1(Q^2)+F_2(Q^2)\big) \gamma_\mu 
+F_2(Q^2)\, \frac{(2p+q)_\mu}{2 m_{B^{(\prime)}_{bc}}}\Bigg]\nu(p),
\end{eqnarray}
where $\nu(p)$ is the Dirac spinor,  $\lambda_{{B^{(\prime)}_{bc}}}$ is the residue of the corresponding baryon; and $F_{1,2}(Q^2)$ are Lorentz invariant form factors whose values at $Q^2 =0$ are needed in determination of the magnetic dipole moments. Summation over spins of $B^{(\prime)}_{bc}$ baryon is carried out as:
\begin{equation}
\label{edmn04}
 \sum_s \nu(p,s)\bar \nu(u,s)=\pslash+m_{B^{(\prime)}_{bc}},
\end{equation}

Using Eqs. (\ref{edmn02})-(\ref{edmn04}) the hadronic representation of the correlation function takes the form
\begin{align}
\label{edmn05}
\Pi^{Had}(p,q)=&\frac{\lambda^2_{B^{(\prime)}_{bc}}}{[(p+q)^2-m^2_{B^{(\prime)}_{bc}}][p^2-m^2_{B^{(\prime)}_{bc}}]}
  \bigg[\Big(F_1(Q^2)
  +F_2(Q^2)\Big)\Big(2\,\pslash\eslash\pslash+\pslash\eslash\qslash+m_{B^{(\prime)}_{bc}}\,\pslash\eslash
  \nonumber\\
  &+
  2\,m_{B^{(\prime)}_{bc}}\,\eslash\pslash+m_{B^{(\prime)}_{bc}}\,\eslash\qslash+m_{B^{(\prime)}_{bc}}^2\eslash
  \Big)
  + \mbox{ other structures~proportional with the} 
  F_2(Q^2) \bigg].
\end{align}
At the static limit, namely $Q^2=0$, the magnetic dipole moment is expressed with the help of the  form factors $F_1(Q^2=0)$ and $F_2(Q^2=0)$ as follows
\begin{equation}
\label{edmn06}
 \mu_{B^{(\prime)}_{bc}}= F_1(Q^2=0)+F_2(Q^2=0).
\end{equation}

We observe that Eq. (\ref{edmn05}) includes different Lorentz structures. To identify the magnetic dipole moment of bottom-charm baryons from Eq. (\ref{edmn05}) we choose the structure $\eslash\qslash$. The reason why we give priority to this structure is that it includes more powers of momentum, which shows the best convergence of the operator product expansion, and therefore causes a more reliable determination of the magnetic dipole moment of bottom-charm baryons.  
As a result, the hadronic representation of the correlation function, the magnetic dipole moment for spin-1/2 $B^{(\prime)}_{bc}$ baryons can be written as:
\begin{eqnarray}
\label{edmn07}
\Pi^{Had}(p,q)&=&\mu_{B^{(\prime)}_{bc}}\frac{\lambda^2_{B^{(\prime)}_{bc}}\,m_{B^{(\prime)}_{bc}}}{[(p+q)^2-m^2_{B^{(\prime)}_{bc}}][p^2-m^2_{B^{(\prime)}_{bc}}]}.
\end{eqnarray}

The second representation of the correlation function, the QCD side, is obtained by explicit use of the interpolating currents to the correlation functions. Then the corresponding quark fields are contracted via  Wick's theorem and the desired results are obtained. After performing the above manipulations, the QCD side of the correlation functions is gained as:
\begin{align}
\label{edmn008}
\Pi_1^{QCD}(p,q)&=i \, \varepsilon^{abc}\varepsilon^{a^{\prime}b^{\prime}c^{\prime}} \int d^4\,x e^{ip\cdot x}
 \langle 0\mid  
 Tr[\gamma_\mu  S_c^{bb^{\prime}}(x)\gamma_\nu  
 \tilde S_b^{aa^{\prime}}(x)] \big( \gamma_5 \gamma^\mu S_q^{cc^{\prime}}(x) \gamma^\nu \gamma_5\big)
 \mid 0\rangle_\gamma, \\
\label{edmn08}
\Pi_2^{QCD}(p,q)&=i \, \varepsilon^{abc}\varepsilon^{a^{\prime}b^{\prime}c^{\prime}} \int d^4\,x e^{ip\cdot x}
 \langle 0\mid  
 Tr[\gamma_5  S_c^{bb^{\prime}}(x)\gamma_5 
 \tilde S_b^{aa^{\prime}}(x)]  S_q^{cc^{\prime}}(x) 
 \mid 0\rangle_\gamma,
\end{align}
where $\tilde S_{Q(q)}^{ij}(x) = CS_{Q(q)}^{{ij}^T}(x)C$ and, $S_q^{ij}(x)$ and
$S_Q^{ij}(x)$ are the light and heavy quark propagators, respectively.
The light and heavy quark propagators are written as~\cite{Yang:1993bp,Belyaev:1985wza},
\begin{align}
\label{edmn09}
S_{q}(x)&=
\frac{1}{2 \pi^2 x^2}\Big( i \frac{{\xslash}}{x^{2}}-\frac{m_{q}}{2 } \Big)
- \frac{ \bar qq }{12} \Big(1-i\frac{m_{q} \xslash}{4}   \Big) 
- \frac{\bar q \sigma.G q }{192}x^2  \Big(1-i\frac{m_{q} \xslash}{6}   \Big)
-\frac {i g_s }{32 \pi^2 x^2} ~G^{\mu \nu} (x) \bigg[\rlap/{x}
\sigma_{\mu \nu} 
+  \sigma_{\mu \nu} \rlap/{x}
 \bigg],
\end{align}%
\begin{eqnarray}
\label{edmn10}
S_{Q}(x)&=&\frac{m_{Q}^{2}}{4 \pi^{2}} \bigg[ \frac{K_{1}(m_{Q}\sqrt{-x^{2}}) }{\sqrt{-x^{2}}}
+i\frac{{\xslash}~K_{2}( m_{Q}\sqrt{-x^{2}})}
{(\sqrt{-x^{2}})^{2}}\bigg]
-\frac{g_{s}m_{Q}}{16\pi ^{2}} \int_{0}^{1}dv~G^{\mu \nu }(vx)\bigg[ \big(\sigma _{\mu \nu }{\xslash}
  +{\xslash}\sigma _{\mu \nu }\big) \frac{K_{1}( m_{Q}\sqrt{-x^{2}}) }{\sqrt{-x^{2}}}\nonumber\\
&& 
+2\sigma ^{\mu \nu }K_{0}( m_{Q}\sqrt{-x^{2}})\bigg],
\end{eqnarray}%
where $G^{\mu \nu }$ is the gluon field strength tensor, $K_{0,1,2}$ are the second kind of Bessel functions;  and $m_q$ and $m_Q$ are the masses of the light and heavy quarks, respectively.

  The correlation functions contain two parts, that is, photon interacting with light and heavy quarks perturbatively (short-distance contributions), and photon interacting with light-quark nonperturbatively (long-distance contributions).  When the photon interacts with light and heavy quarks perturbatively, one of the propagators in Eq. (\ref{edmn08}) is substituted by
 \begin{eqnarray}
\label{edmn12}
S^{free} (x) \rightarrow \int d^4y\, S^{free} (x-y)\,\rlap/{\!A}(y)\, S^{free} (y)\,,
\end{eqnarray}
where $S^{free} (x)$ is the first term of the light or heavy quark propagators and the remaining two quark propagators are taken as full quark propagators. When the manipulations mentioned above are performed, the $S_q^{free} (x)$ and $S_Q^{free} (x)$ take the following forms
\begin{eqnarray}\label{sfreepert}
&& S_q^{free}(x)=\frac{e_q}{32 \pi^2 x^2}\Big(\varepsilon_\alpha q_\beta-\varepsilon_\beta q_\alpha\Big)
 \Big(\xslash\sigma_{\alpha \beta}+\sigma_{\alpha\beta}\xslash\Big),\nonumber\\
&& S_Q^{free}(x)=-i\frac{e_Q m_Q}{32 \pi^2}
\Big(\varepsilon_\alpha q_\beta-\varepsilon_\beta q_\alpha\Big)
\Big[2\sigma_{\alpha\beta}K_{0}\Big( m_{Q}\sqrt{-x^{2}}\Big)
 +\frac{K_{1}\Big( m_{Q}\sqrt{-x^{2}}\Big) }{\sqrt{-x^{2}}}
 \Big(\xslash\sigma_{\alpha \beta}+\sigma_{\alpha\beta}\xslash\Big)\Big].
\end{eqnarray}

Eq. (\ref{sfreepert}) is inserted into Eqs. (\ref{edmn008}) and (\ref{edmn08}), and as a result of these calculations for the perturbative contributions we obtain  
\begin{align}
\label{pert1}
\Pi_{1-Pert}^{QCD}(p,q)&=i \, \varepsilon^{abc}\varepsilon^{a^{\prime}b^{\prime}c^{\prime}} \int d^4\,x e^{ip\cdot x}
 \Big\{  
 Tr[\gamma_\mu  S_c^{free}(x)\gamma_\nu  \tilde S_b^{aa^{\prime}}(x)] \big( \gamma_5 \gamma^\mu S_q^{cc^{\prime}}(x) \gamma^\nu \gamma_5\big) \delta^{bb^{\prime}} \nonumber\\
 &
 +Tr[\gamma_\mu  S_c^{bb^{\prime}}(x)\gamma_\nu  \tilde S_b^{free}(x)] \big( \gamma_5 \gamma^\mu S_q^{cc^{\prime}}(x) \gamma^\nu \gamma_5\big) \delta^{aa^{\prime}} \nonumber\\
 &+Tr[\gamma_\mu  S_c^{bb^{\prime}}(x)\gamma_\nu  \tilde S_b^{aa^{\prime}}(x)] \big( \gamma_5 \gamma^\mu S_q^{free}(x) \gamma^\nu \gamma_5\big) \delta^{cc^{\prime}}
 \Big\},\\
 \nonumber\\
\label{pert2}
\Pi_{2-Pert}^{QCD}(p,q)&=i \, \varepsilon^{abc}\varepsilon^{a^{\prime}b^{\prime}c^{\prime}} \int d^4\,x e^{ip\cdot x}
 \Big \{
 Tr[\gamma_5  S_c^{free}(x)\gamma_5  \tilde S_b^{aa^{\prime}}(x)]  S_q^{cc^{\prime}}(x)  \delta^{bb^{\prime}}\nonumber\\ 
 &+  Tr[\gamma_5  S_c^{bb^{\prime}}(x)\gamma_5  \tilde S_b^{free}(x)]  S_q^{cc^{\prime}}(x)  \delta^{aa^{\prime}}\nonumber\\
 &+
 Tr[\gamma_5  S_c^{bb^{\prime}}(x)\gamma_5  \tilde S_b^{aa^{\prime}}(x)]  S_q^{free}(x)  \delta^{cc^{\prime}}
 \Big\}.
\end{align}

Note that all possibilities are considered in the above equations. In the first line of  Eqs. (\ref{pert1}) and (\ref{pert2}), the photon interacts perturbatively with the heavy quark propagator, while the remaining two propagators are considered full. Likewise, in the second line of Eqs. (\ref{pert1}) and (\ref{pert2}), the photon interacts perturbatively with the heavy quark propagators, while the other propagators are assumed to be full, etc.

In the case of the photon interacting with light quark nonperturbatively in Eqs. (\ref{edmn008}) and (\ref{edmn08}) is substituted by
\begin{align}
\label{edmn13}
S_{\alpha\beta}^{ab} \rightarrow -\frac{1}{4} (\bar{q}^a \Gamma_i q^b)(\Gamma_i)_{\alpha\beta},
\end{align}
where $\Gamma_i = I, \gamma_5, \gamma_\mu, i\gamma_5 \gamma_\mu, \sigma_{\mu\nu}/2$ and the remaining two propagators are substituted with the full quark propagators.  In this case, the correlation functions take the form,
\begin{align}
\label{nonpert1}
\Pi_{1-Nonpert}^{QCD}(p,q)&=-\frac{i}{4} \, \varepsilon^{abc}\varepsilon^{a^{\prime}b^{\prime}c^{\prime}} \int d^4\,x e^{ip\cdot x}
 \langle 0\mid  
 Tr[\gamma_\mu  S_c^{bb^{\prime}}(x)\gamma_\nu   \tilde S_b^{aa^{\prime}}(x)] \big( \gamma_5 \gamma^\mu \Gamma_i  \gamma^\nu \gamma_5\big) \big(\bar q ^c (x) \Gamma_i  q^{c^{\prime }}(0)\big)
 \mid 0\rangle_\gamma, \\
\label{nonpert2}
\Pi_{2-Nonpert}^{QCD}(p,q)&=-\frac{i}{4} \, \varepsilon^{abc}\varepsilon^{a^{\prime}b^{\prime}c^{\prime}} \int d^4\,x e^{ip\cdot x}
 \langle 0\mid   Tr[\gamma_5  S_c^{bb^{\prime}}(x)\gamma_5   \tilde S_b^{aa^{\prime}}(x)]  \,\Gamma_i \,  \big(\bar q ^c (x) \Gamma_i  q^{c^{\prime }}(0)\big)
 \mid 0\rangle_\gamma,
\end{align}

 By replacing the light quark propagators and using the expression $  \bar q^a(x)\Gamma_i q^{a'}(0)\rightarrow \frac{1}{3}\delta^{aa'}\bar q(x)\Gamma_i q(0)$,  Eq. (\ref{nonpert1}) and Eq. (\ref{nonpert2})  takes the form
\begin{align}
\label{nonpert3}
\Pi_{1-Nonpert}^{QCD}(p,q)&=-i \, \varepsilon^{abc}\varepsilon^{a^{\prime}b^{\prime}c^{\prime}} \int d^4\,x e^{ip\cdot x}
  Tr[\gamma_\mu  S_c^{bb^{\prime}}(x)\gamma_\nu   \tilde S_b^{aa^{\prime}}(x)] \big( \gamma_5 \gamma^\mu \Gamma_i  \gamma^\nu \gamma_5\big) 
  \frac{1}{12} \langle \gamma(q) |\bar q(x)\Gamma_i q(0)|0\rangle , \\
\label{nonpert4}
\Pi_{2-Nonpert}^{QCD}(p,q)&=-i \, \varepsilon^{abc}\varepsilon^{a^{\prime}b^{\prime}c^{\prime}} \int d^4\,x e^{ip\cdot x}
  Tr[\gamma_5  S_c^{bb^{\prime}}(x)\gamma_5   \tilde S_b^{aa^{\prime}}(x)]  \,\Gamma_i \,  
  \frac{1}{12} \langle \gamma(q) |\bar q(x)\Gamma_i q(0)|0\rangle .
\end{align}

Moreover, if a light quark interacts nonperturbatively with a photon, a gluon can also be released from one of the remaining two quark propagators.  Matrix elements such as $\langle \gamma(q) |\bar q(x)\Gamma_i G_{\alpha\beta}(vx) q(0)|0\rangle $ appear when the calculations necessary to include these contributions are performed. 
To evaluate the nonperturbative effects, we require $\langle \gamma(q)\vel \bar{q}(x) \Gamma_i q(0) \ver 0\rangle$ and $\langle \gamma(q)\vel \bar{q}(x) \Gamma_i G_{\alpha\beta}q(0) \ver 0\rangle$ matrix elements and those are determined in connection with the photon distribution amplitude (DAs) with definite twists in  Ref.~\cite{Ball:2002ps}.  
 It is worth noting that the photon DAs used in this study include contributions from only light quarks. However, principally, the photon can be emitted at a long-distance from the heavy quarks. 
Technically speaking, the matrix elements of nonlocal operators are proportional to the product of DAs, quark condensates, and some nonperturbative constants.  Since we know that the contribution of non-perturbative constants to our analysis is negligible, even in the case of light quarks, we can neglect these contributions in the case of heavy quarks. The heavy quark condensates are proportional to $1/m_Q$. Due to the large mass of the heavy quarks, such condensates for the heavy quarks will be largely suppressed~\cite{Antonov:2012ud}.  Therefore, DAs containing heavy quarks (long-distance contributions) were not used in our computations. Only the short-distance photon emission from the heavy quarks is considered, as described in Eq. (\ref{edmn12}). 
 The QCD representation of the correlation function can be evaluated in connection with QCD parameters by replacing photon DAs and expressions for heavy and light quarks propagators into Eq. (\ref{edmn08}).

The sum rules are acquired by equating the expression of the correlation function in connection with QCD parameters to its expression in connection with the hadron features, employing their spectral representation. To suppress the subtraction terms in the spectral representation of the correlation function, the double Borel transformation in connection with the variables $p^2$ and $(p + q)^2$ is applied. After the transformation, contributions from the higher states and continuum are also exponentially suppressed. Finally, we pick the  $\eslash\qslash$ structure for the magnetic dipole moments and obtain
\begin{eqnarray}
\label{edmn14}
 \mu_{B_{bc}}&=&\frac{e^{\frac{m^2_{B_{bc}}}{M^2}}}{\lambda^2_{_{bc}}\, m_{B_{bc}}}\, \Pi_1^{QCD}(M^2,s_0),\\
 \mu_{B^{\prime}_{bc}}&=&\frac{e^{\frac{m^2_{B^{\prime}_{bc}}}{M^2}}}{\lambda^2_{B^{\prime}_{bc}}\, m_{B^{\prime}_{bc}}}\, \Pi_2^{QCD}(M^2,s_0).\label{edmn15}
\end{eqnarray}
%
The analytical expressions of the functions $\Pi_1^{QCD}(M^2,s_0)$ and $\Pi_2^{QCD}(M^2,s_0)$ are very similar forms, so, for the sake of brevity, we present only the explicit expressions of the function $\Pi_1^{QCD}(M^2,s_0)$, which is given in appendix \ref{appendpi}.

 \end{widetext}

\section{Numerical analysis and conclusion}\label{numerical}

In this section, numerical calculations for the spin-$\frac{1}{2}$ bottom-charm baryons have been presented.  The numerical values of the parameters used in this section of the analysis are as follows:  $m_u =m_d =0$,
$m_s=93.4^{+8.6}_{-3.4}$~MeV,
$m_c = 1.27 \pm 0.02$~GeV, $m_b=4.18^{+0.03}_{-0.02}$~GeV,~\cite{ParticleDataGroup:2022pth},  
$\langle \bar qq\rangle =(-0.24\pm0.01)^3$~GeV$^3$ \cite{Ioffe:2005ym},
$m_0^{2} = 0.8 \pm 0.12$~GeV$^2$, $\langle g_s^2G^2\rangle = 0.88~ GeV^4$~\cite{Matheus:2006xi}
 and $\chi=-2.85 \pm 0.5$~GeV$^2$~\cite{Rohrwild:2007yt}. The masses  and residues of the bottom-charm 
 baryons are borrowed from Ref.~\cite{Wang:2022ufh}.
Another set of main input parameters are the photon wavefunctions of different twists, entering the DAs. These wavefunctions are given in appendix \ref{photonDAs}. 

\begin{widetext}
 
\begin{figure}
\centering
 \subfloat[]{\includegraphics[width=0.4\textwidth]{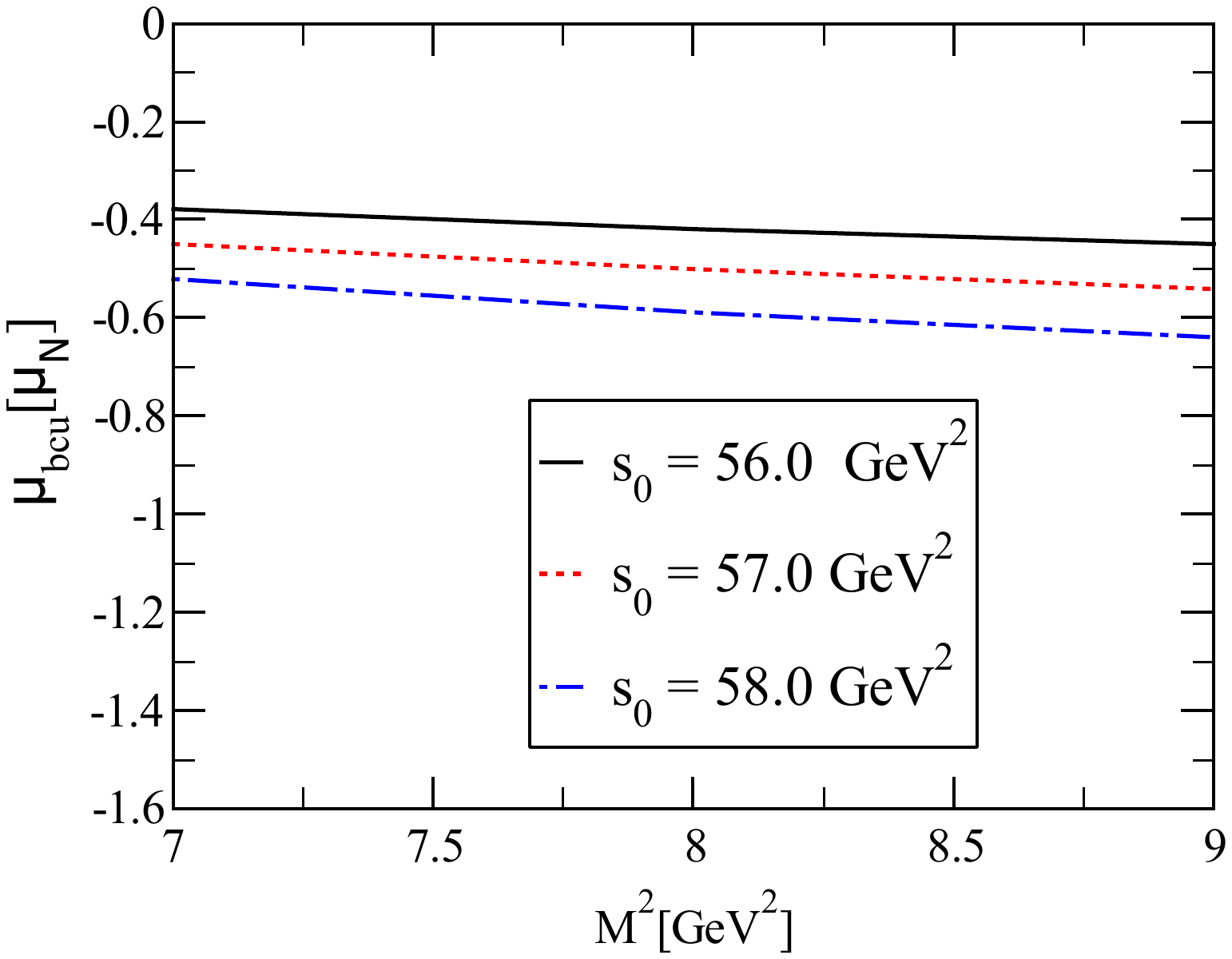}}~~~~
 \subfloat[]{ \includegraphics[width=0.4\textwidth]{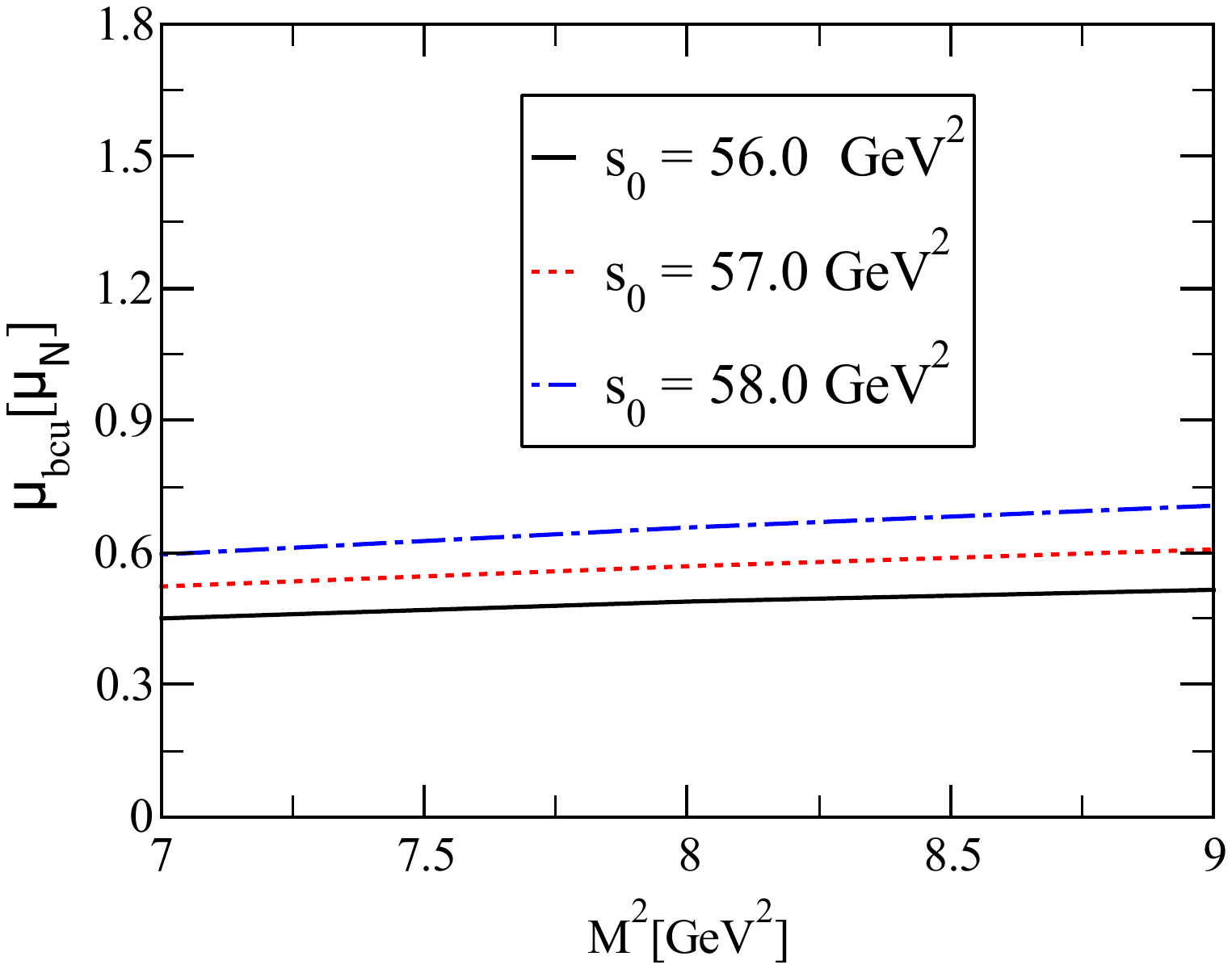}}\\
  \vspace{0.5cm}
 \subfloat[]{\includegraphics[width=0.4\textwidth]{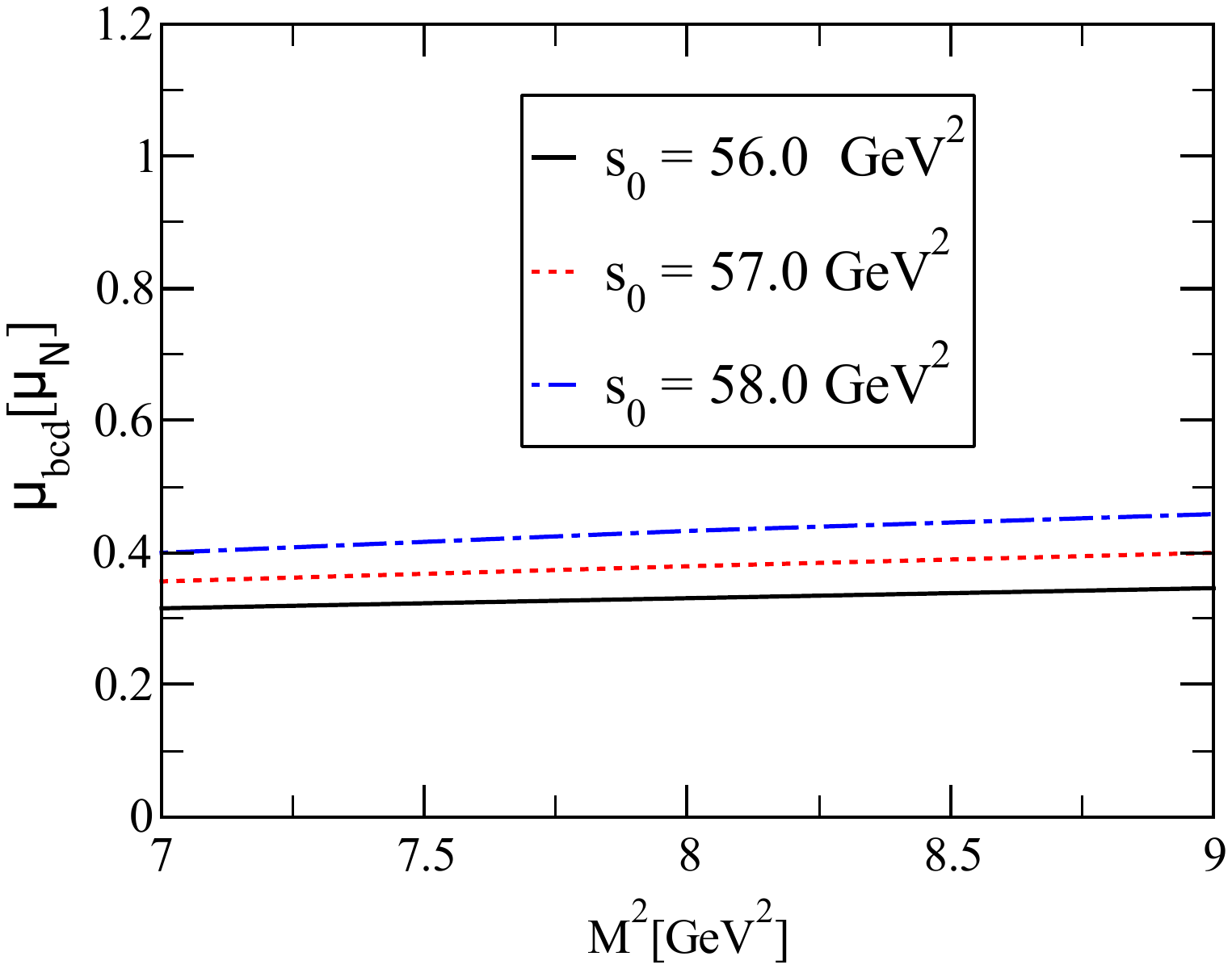}}~~~~
 \subfloat[]{ \includegraphics[width=0.4\textwidth]{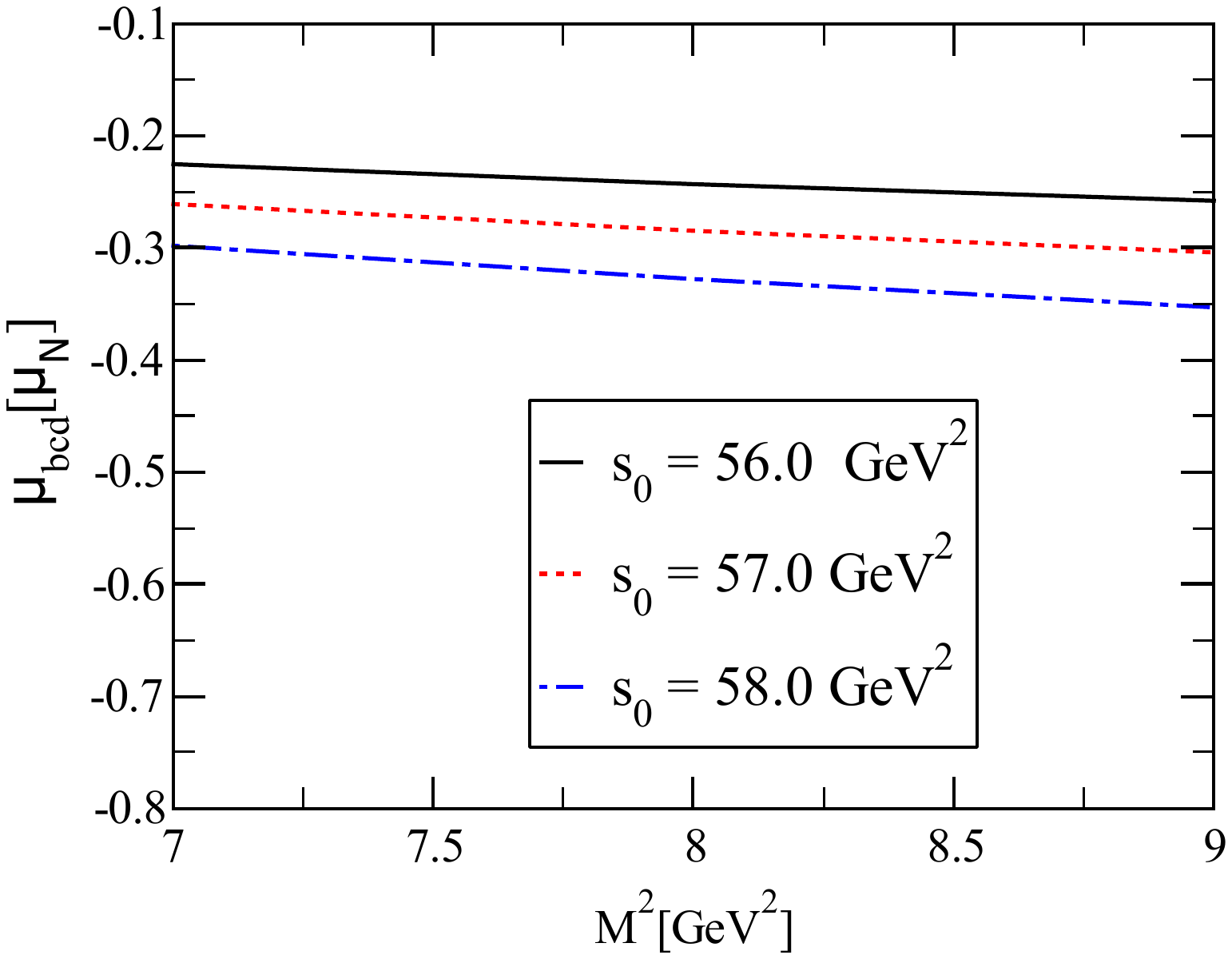}}\\
  \vspace{0.5cm}
 \subfloat[]{\includegraphics[width=0.4\textwidth]{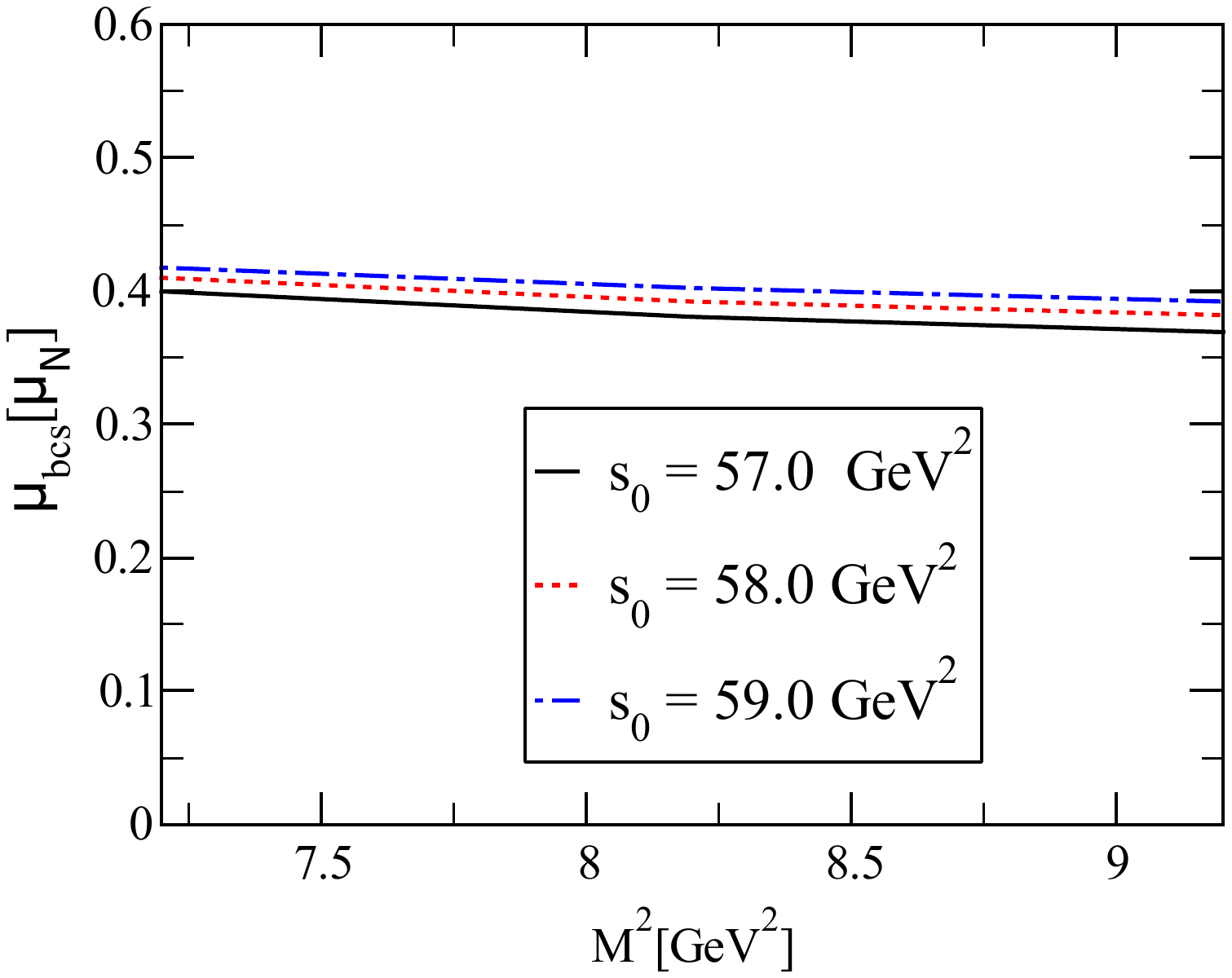}}~~~~
 \subfloat[]{ \includegraphics[width=0.4\textwidth]{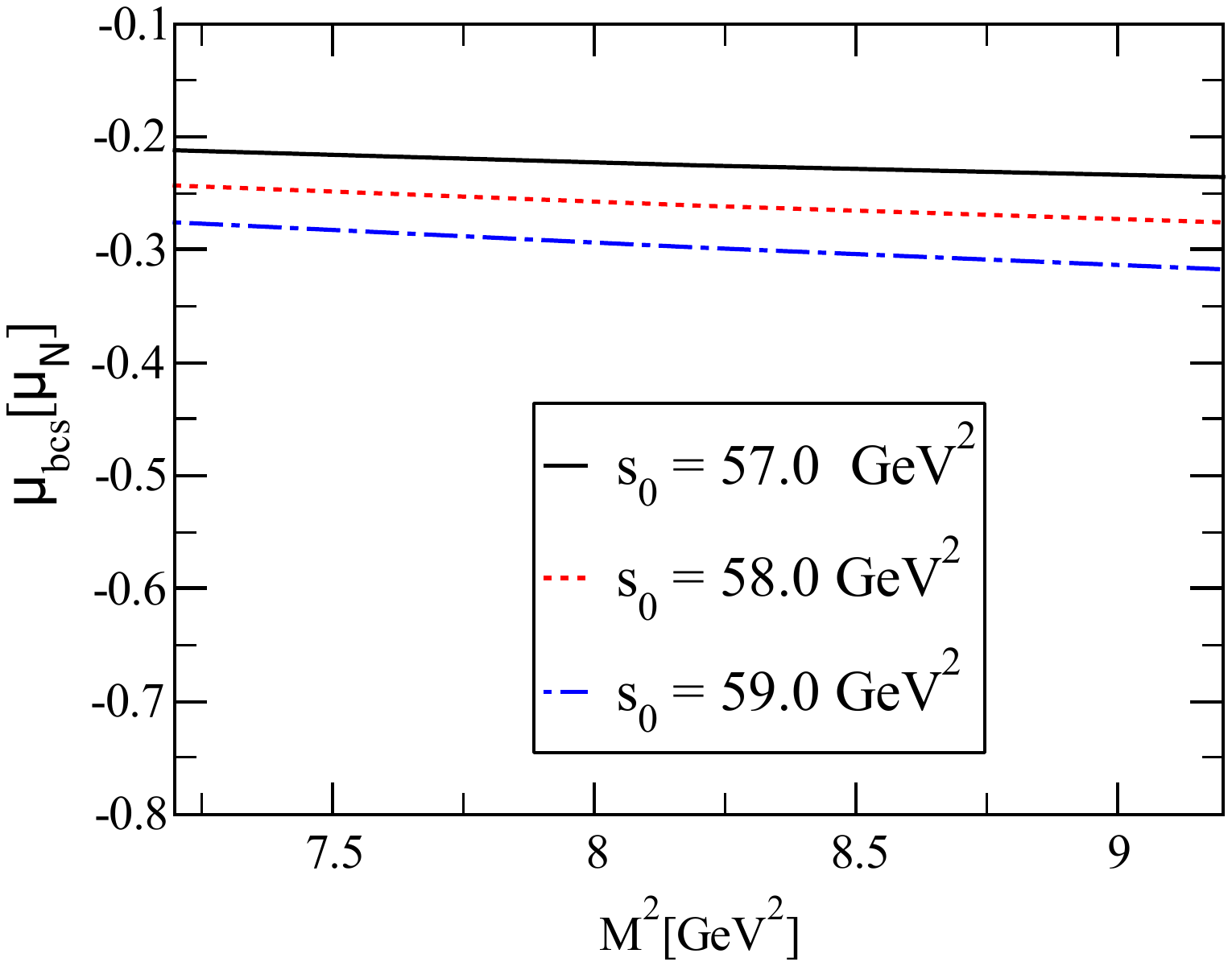}}
 \caption{ The magnetic dipole moments of the bottom-charm baryons with variations of Borel parameter $M^{2}$
   at the different values of the continuum threshold parameter $s_0$: 
 (a), (c) and (e) for the $\Xi_{bc}^{+} $, $\Xi_{bc}^{0}$ and $\Omega_{bc}^{0}$ baryons,
 (b), (d) and (f) for the $\Xi_{bc}^{ \prime +} $, $\Xi_{bc}^{\prime 0}$ and $\Omega_{bc}^{\prime 0}$ baryons.}
 \label{Msqfig}
  \end{figure}
  
  \end{widetext}

As one can see from Eqs. (\ref{edmn14}) and (\ref{edmn15}), the magnetic dipole moments of the bottom-charm baryons, in addition to the above-mentioned input parameters, contain also two helping parameters:  the continuum threshold $s_0$ and the Borel mass parameter $M^2$. 
  To carry out numerical computations, it is also necessary to choose the working intervals for the parameters $s_0$ and $M^2$.  
We apply the OPE convergence and pole dominance conditions to determine the working intervals of $s_0$ and $M^2$.
 From this perspective, we determine the value of these helping parameters within the interval
 $ 56.0$~GeV$^2 \leq s_0 \leq 58.0 $~GeV$^2$ and $7.0$~GeV$^2$ $\leq$ M$^2$ $\leq$ $9.0$~GeV$^2$  for $\Xi_{bc}^{(\prime)}$ baryons; and $ 57.0$~GeV$^2 \leq s_0 \leq 59.0 $~GeV$^2$ and $7.2$~GeV$^2$ $\leq$ M$^2$ $\leq$ $9.2$~GeV$^2$ for $\Omega_{bc}^{(\prime)}$ baryon.
By using all the inputs as well as the working intervals of the helping parameters, we plot the variation of the magnetic dipole moments concerning the helping parameters for the considered structure in Fig. \ref{Msqfig}. This figure indicates the mild dependence of the magnetic dipole moments on the variations of the helping parameters in their working intervals.

The obtained magnetic dipole moment results are given in Table I, and the same table also shows the predictions of other theoretical models for the magnetic dipole moments of bottom-charm baryons are also presented. The uncertainties in the values of the input parameters and photon DAs, as well as the variations in the calculations of the working windows $s_0$ and $M^2$, are responsible for the reported errors in the results. As can be seen in Table I, the magnetic dipole moments obtained for $\Xi_{bc}$ and $\Omega_{bc}$ baryons have opposite signs to those obtained for $\Xi^{\prime}_{bc}$ and $\Omega^{\prime}_{bc}$ baryons.  The main reason for this is that the structure of these baryons is symmetric and antisymmetric under the exchange of $b$- and $c$-quarks.
We have also been able to determine individual quark contributions to the magnetic dipole moments. 
Since the contribution of the heavy quarks to the magnetic dipole moments is proportional to $1/m_Q$, the main contribution is expected to come from the light-quark. In our calculations, the light-quark comes through the two heavy quarks and gives the dominant contribution.
When analyzing the interpolating current couple to the $\Xi^{\prime}_{bc}$ and $\Omega^{\prime}_{bc}$ baryons, it can be seen that the heavy diquark has a scalar structure. Therefore, the contribution of the heavy diquark to the magnetic dipole moment is expected to be either very small or not at all. We find that the magnetic dipole moments of these baryons are due almost entirely to the light quarks.  Our final comment on the magnetic dipole moment results is an examination of the violation of the $U$-symmetry. Though the $U$-symmetry breaking effects have been considered via a nonzero s-quark mass and s-quark condensate, we observe that the $U$- symmetry violation in the magnetic dipole moments is quite small for both symmetric and antisymmetric interpolating currents.

Magnetic dipole moments of  $\Xi_{bc}^{(\prime)}$ and $\Omega_{bc}^{(\prime)}$ baryons 
have been previously investigated in the nonrelativistic quark model (NQM)~\cite{Silvestre-Brac:1996myf,Albertus:2006ya}, relativistic harmonic confinement model (RHM)~\cite{Gadaria:2016omw},  effective quark mass (EQM) and shielded quark charge scheme (SQCS)~\cite {Dhir:2013nka}, MIT bag model~\cite{Simonis:2018rld, Bernotas:2012nz},  hypercentral constituent quark model (HCQM)~\cite{Shah:2016vmd,Shah:2017liu},  heavy baryon chiral perturbation theory (HB$\chi$PT)~\cite{Li:2017cfz} and relativistic quark model (RQM)\cite{Faessler:2006ft}. It is seen that the signs of the magnetic dipole moments are correctly determined. However, different theoretical models have yielded quite different results for the magnetic dipole moments of bottom-charm baryons, which can be used to distinguish between models. The choice of wave functions in different models may be responsible for these discrepancies. However, the origin of this obvious discrepancy remains an open issue. It is clear that further theoretical and experimental studies are required to clarify these inconsistencies and to better understand the current situation. However, direct measurements of the magnetic dipole moments of bottom-charm baryons are not yet possible. Therefore, any indirect projections of the magnetic dipole moments of the bottom-charm baryons would be quite helpful.

\begin{widetext}

  \begin{table}[htp]\label{comp}
 \addtolength{\tabcolsep}{10pt}
   \centering
\begin{tabular}{lccccccccc}
\hline\hline
\\
~Models ~&~ $\mu_{\Xi_{bc}^{+}} $ ~&~ $\mu_{\Xi_{bc}^{0}} $~ &~ $\mu_{\Omega_{bc}^{0}} $
&~ $\mu_{\Xi_{bc}^{\prime +}} $ ~&~ $\mu_{\Xi_{bc}^{\prime 0}} $~ &~ $\mu_{\Omega_{bc}^{\prime 0}}$\\
\\
\hline\hline
\\
NQM \cite{Silvestre-Brac:1996myf} &-0.198  & 0.058 & 0.009 &- &- &-\\ 
\\
NQM \cite{Albertus:2006ya} &-0.475 & 0.518 & 0.368 &1.99 & -0.993 &-0.542\\
\\
RHM \cite{Gadaria:2016omw}& -0.52 & 0.63 & 0.49 &- &- &-\\
\\
EQM~\cite {Dhir:2013nka}& -0.387 & 0.499 & 0.399 &1.729 &-0.864 &-0.580\\
\\
SCQS~\cite {Dhir:2013nka}& -0.369 & 0.48 & 0.407 &1.718 &-0.817 & -0.621\\
\\
Bag model \cite{Simonis:2018rld} & -0.35 & 0.446 & 0.378 & 1.59 & -0.796 & -0.595\\
\\
Bag model \cite{Bernotas:2012nz} & -0.25 & 0.13 & 0.08 & 1.71 &-0.53 & -0.27\\
\\
HCQM~\cite{Shah:2016vmd,Shah:2017liu} &-0.204 & 0.354 & 0.439 &- &- &-\\
\\
HB$\chi$PT~\cite{Li:2017cfz} & -0.54 & 0.56 &0.49&0.69 &-0.59 &0.24\\
\\
RQM \cite{Faessler:2006ft} &-0.12 & 0.42 &0.45 & 1.52 &-0.76 &- 0.61 \\
\\
This work&$-0.50^{+0.14}_{-0.12}$&$0.39^{+0.06}_{-0.05}$&$0.38^{+0.05}_{-0.04}$
&$0.57^{+0.13}_{-0.12}$&$-0.29^{+0.07}_{-0.06}$&$-0.26^{+0.06}_{-0.05}$\\
\\
\hline\hline
\end{tabular}
\caption{Comparison of magnetic dipole moments of bottom-charm baryons in the literature, including the nonrelativistic quark model (NQM)~\cite{Silvestre-Brac:1996myf,Albertus:2006ya}, relativistic harmonic confinement model (RHM)~\cite{Gadaria:2016omw},  effective quark mass (EQM) and shielded quark charge scheme (SQCS)~\cite {Dhir:2013nka}, MIT bag model~\cite{Simonis:2018rld, Bernotas:2012nz},  hypercentral constituent quark model (HCQM)~\cite{Shah:2016vmd,Shah:2017liu},  heavy baryon chiral perturbation theory (HB$\chi$PT)~\cite{Li:2017cfz} and relativistic quark model (RQM)\cite{Faessler:2006ft} (in nuclear magneton $\mu_N$).}  
 \end{table}

 \end{widetext}

\section{Summary}\label{summary} 

The observation of the $\Xi^{++}_{cc}$ baryon by the LHCb Collaboration has aroused great interest in doubly-heavy baryon systems. The measurement of the features of doubly-heavy baryons ensures insight into both their inner structure and production mechanism. In this work, we have studied the magnetic dipole moments of the bottom-charm baryons within the framework of QCD light-cone sum rules.  We have compared our results with other theoretical predictions that could be a useful complementary tool for the interpretation of the doubly-heavy baryon sector, and we observe that they are not in mutual agreement with each other. Different theoretical models give quite different predictions for the magnetic dipole moments of the bottom-charm baryons, as can be seen from the results in Table I.  Comparison of the acquired results on the magnetic dipole moments of the $\Omega^{(\prime)0}_{bc}$ baryon with those of $\Xi^{(\prime)0}_{bc}$ baryon presents a small $U$-symmetry violation.  The magnetic dipole moment is the leading-order response of a bound system to a weak external magnetic field and thus provides an excellent platform to probe the internal organization of hadrons, which is governed by the quark-gluon dynamics of QCD. Moreover, understanding the electromagnetic properties of doubly-heavy baryons is the key to confinement and heavy flavor effects. Future experimental efforts on the properties of doubly-heavy baryons may figure out inconsistencies among different model estimations. We believe that our results will be helpful in future experimental and theoretical attempts regarding doubly-heavy baryons.

\section{Acknowledgements}
The author would like to thank A. \"{O}zpineci for useful discussions, comments, and remarks.

  \begin{widetext}
   
   \appendix
  \section{The explicit form of the $\Pi_1^{QCD}(M^2,s_0)$ function} \label{appendpi}
  
  In this appendix, we present the explicit expression for the $\Pi_1^{QCD}(M^2,s_0)$ function:
 \begin{align}
 \Pi_1^{QCD}(M^2,s_0) &= \frac{e_q P_1 P_2}{1327104 \pi} \Big[-6  \Big (I[0, 2, 2, 0] - 2 I[0, 2, 3, 0] + I[0, 2, 4, 0] + 
      2 I[1, 1, 2, 0] - 4 I[1, 1, 3, 0] + 2 I[1, 1, 4, 0]\Big) \nonumber\\
      &\times I_2[ h_\gamma] + 
   3  \Big (5 I[0, 2, 2, 0] - 10 I[0, 2, 3, 0] + 5 I[0, 2, 4, 0] + 
      8 I[1, 1, 2, 0] - 16 I[1, 1, 3, 0] + 8 I[1, 1, 4, 0]\Big) \nonumber\\
      &\times \mathbb A[u_ 0] + 2 \chi \Big (4 I[0, 3, 2, 0] - 8 I[0, 3, 3, 0] + 
       4 I[0, 3, 4, 0] + 3 I[1, 2, 2, 0] - 6 I[1, 2, 3, 0] + 
       3 I[1, 2, 4, 0]\Big) \nonumber\\
      &\times \varphi_\gamma[u_ 0]\Big]\nonumber\\
      &-\frac{P_2}{983040 \pi} \Big[5 e_q \Big (16 m_c^2  \big (I[0, 3, 1, 0] - 
      I[0, 3, 2, 0]\big) I_ 1[\mathcal {S}] + 
   3\big (3 I[0, 4, 2, 0] - 6 I[0, 4, 3, 0] + 3 I[0, 4, 4, 0]\nonumber\\
      & + 
      4 I[1, 3, 2, 0] - 8 I[1, 3, 3, 0] + 
      4 I[1, 3, 4, 0]\big)  I_1[\mathcal {S}] + 
   9  \big (-I[0, 4, 3, 0] + 3 I[0, 4, 4, 0] - 3 I[0, 4, 5, 0] \nonumber\\
      &+ 
      I[0, 4, 6, 0] - 4 I[1, 3, 3, 0] + 12 I[1, 3, 4, 0] - 
      12 I[1, 3, 5, 0] + 4 I[1, 3, 6, 0]\big) I_ 2[h_\gamma] + 
   9  \big (5 I[0, 4, 3, 0] \nonumber\\
      &- 15 I[0, 4, 4, 0] + 15 I[0, 4, 5, 0] - 
      5 I[0, 4, 6, 0] + 16 I[1, 3, 3, 0] - 48 I[1, 3, 4, 0] + 
      48 I[1, 3, 5, 0]\nonumber\\
      & - 16 I[1, 3, 6, 0] + 6 I[2, 2, 3, 0] - 
      18 I[2, 2, 4, 0] + 18 I[2, 2, 5, 0] - 6 I[2, 2, 6, 0]\big) \mathbb A[
      u_ 0] +
      9 \chi  \big(4 I[0, 5, 3, 0] \nonumber\\
      &- 12I[0, 5, 4, 0] + 12 I[0, 5, 5, 0] - 
   4 I[0, 5, 6, 0] + 5 I[1, 4, 3, 0] - 15 I[1, 4, 4, 0] + 
   15 I[1, 4, 5, 0] \nonumber\\
      & - 5 I[1, 4, 6, 0]\big) \varphi_\gamma[u_0]\Big)
      +e_b \Big (32 m_b m_c \big (I[0, 3, 1, 0] - 2 I[0, 3, 2, 0] + 
      I[0, 3, 3, 0]\big) + 18 I[0, 4, 2, 0] \nonumber\\
      &- 54 I[0, 4, 3, 0] + 
   54 I[0, 4, 4, 0] - 18[0, 4, 5, 0] + 
   3 m_ 0^2 \Big (16 m_b m_c \big (I[0, 2, 1, 0] - 2 I[0, 2, 2, 0] + 
          I[0, 2, 3, 0] \nonumber\\
      &+ I[1, 1, 1, 0] - 2 I[1, 1, 2, 0] + 
          I[1, 1, 3, 0]\big) + 12 I[0, 3, 2, 0] - 36 I[0, 3, 3, 0] + 
       36 I[0, 3, 4, 0] - 12 I[0, 3, 5, 0]\nonumber\\
      & + 45 I[1, 2, 2, 0] - 
       135 I[1, 2, 3, 0] + 135 I[1, 2, 4, 0] - 45 I[1, 2, 5, 0] + 
       6 I[2, 1, 2, 0] - 18 I[2, 1, 3, 0] \nonumber\\
      &+ 18 I[2, 1, 4, 0] - 
       6 I[2, 1, 5, 0]\Big)\Big)
       +5 e_c \Big (32 m_b m_c \big (I[0, 3, 2, 0] - I[0, 3, 3, 0]\big) + 
   18  I[0, 4, 3, 0] \nonumber\\
      &- 36 I[0, 4, 4, 0] + 18 I[0, 4, 5, 0] + 
   3 m_ 0^2 \Big (16 m_b m_c \big (I[0, 2, 2, 0] - I[0, 2, 3, 0] + 
          I[1, 1, 2, 0] - I[1, 1, 3, 0]\big) \nonumber\\
      &+ 12 I[0, 3, 3, 0] - 
       24 I[0, 3, 4, 0] + 12 I[0, 3, 5, 0] + 45 I[1, 2, 3, 0] - 
       90 I[1, 2, 4, 0] + 45 I[1, 2, 5, 0]\nonumber\\
      & + 6 I[2, 1, 3, 0] - 
       12 I[2, 1, 4, 0] + 6 I[2, 1, 5, 0]\Big)\Big)
      \Big] \nonumber\\
   &  + \frac{m_q m_c m_b}{1024 \pi^3}\Big[e_c \big (I[0, 4, 2, 0] - I[0, 4, 2, 1] - I[0, 4, 3, 0]\big) + 
 e_b \big (I[0, 4, 1, 0] - 2 I[0, 4, 1, 1] + I[0, 4, 1, 2] \nonumber\\
      &- 
    2 I[0, 4, 2, 0] + 2 I[0, 4, 2, 1] + I[0, 4, 3, 0]\big)
      \big]\nonumber\\
        &+ \frac{3m_q m_0^2}{327680 \pi^3} \Big[e_b \Big (20 I[0, 4, 2, 0] - 60 I[0, 4, 2, 1] + 60 I[0, 4, 2, 2] - 
    20 I[0, 4, 2, 3] - 60 I[0, 4, 3, 0] \nonumber\\
      &+ 120 I[0, 4, 3, 1] - 
    60 I[0, 4, 3, 2] + 60 I[0, 4, 4, 0] - 60 I[0, 4, 4, 1] - 
    20 I[0, 4, 5, 0] - I[0, 5, 2, 0] \nonumber\\    
      &+ 3 I[0, 5, 3, 0] - 
    3 I[0, 5, 4, 0] + I[0, 5, 5, 0] - 
    80 \big (I[1, 3, 2, 0] - 3 I[1, 3, 2, 1] + 3 I[1, 3, 2, 2] - 
        I[1, 3, 2, 3] \nonumber\\
      &- 
        3 (I[1, 3, 3, 0] - 2 I[1, 3, 3, 1] + I[1, 3, 3, 2] - 
           I[1, 3, 4, 0] + I[1, 3, 4, 1]) - I[1, 3, 5, 0]\big)\Big) \nonumber\\
                &+ 
 e_c \Big (20 I[0, 4, 3, 0] - 40 I[0, 4, 3, 1] + 20 I[0, 4, 3, 2] - 
    40 I[0, 4, 4, 0] + 40 I[0, 4, 4, 1] + 20 I[0, 4, 5, 0] \nonumber\\
      &- 
    I[0, 5, 3, 0] + 2 I[0, 5, 4, 0] - I[0, 5, 5, 0] - 
    80 \big (I[1, 3, 3, 0] - 2 I[1, 3, 3, 1] + I[1, 3, 3, 2] - 
        2 I[1, 3, 4, 0] \nonumber\\
      &+ 2 I[1, 3, 4, 1] + I[1, 3, 5, 0]\big)\big)      
      \Big],
\end{align}
where
\begin{align*}
 {M^2}= \frac{M_1^2 M_2^2}{M_1^2+M_2^2}, ~~~
 u_0= \frac{M_1^2}{M_1^2+M_2^2}, 
\end{align*}
with $ M_1^2 $ and $ M_2^2 $ being the Borel parameters in the initial and final states, respectively. We set $ M_1^2= M_2^2 =2 M^2 $  as the initial and final states are the same. We will fix $ M^2 $ and continuum threshold $ s_0 $ based on the standard prescription of the LCSR method in the next section. Here  $P_1 =\langle g_s^2 G^2\rangle$ is gluon condensate,  $\chi$ stands for the magnetic susceptibility of the quark condensate, and $P_2 =\langle \bar q q \rangle$, $e_q$ and $m_q$  are the quark condensate, electric charge and mass of the u, d or s-quark, respectively. 

Explicit forms of ~$I[n,m,l,k]$ and $I_i[\mathcal{F}]$ functions are as follows:
\begin{align}
 I[n,m,l,k]&= \int_{4 m_c^2}^{s_0} ds \int_{0}^1 dt \int_{0}^1 dw~ e^{-s/M^2}~
 s^n\,(s-4\,m_c^2)^m\,t^l\,w^k,\nonumber\\
   I_1[\mathcal{F}]&=\int D_{\alpha_i} \int_0^1 dv~ \mathcal{F}(\alpha_{\bar q},\alpha_q,\alpha_g)
 \delta(\alpha_ q +\bar v \alpha_g-u_0),\nonumber\\
 I_2[\mathcal{F}]&=\int_0^1 du~ \mathcal{F}(u),
 \end{align}
 where $\mathcal{F}$ denotes the corresponding photon DAs.
 


\section{ Distribution Amplitudes of the photon } \label{photonDAs}
In this appendix, the matrix elements $\langle \gamma(q)\vel \bar{q}(x) \Gamma_i q(0) \ver 0\rangle$  
and $\langle \gamma(q)\vel \bar{q}(x) \Gamma_i G_{\mu\nu}q(0) \ver 0\rangle$ associated with the photon DAs are presented as follows \cite{Ball:2002ps}:
\begin{eqnarray*}
\label{esbs14}
&&\langle \gamma(q) \vert  \bar q(x) \gamma_\mu q(0) \vert 0 \rangle
= e_q f_{3 \gamma} \left(\varepsilon_\mu - q_\mu \frac{\varepsilon
x}{q x} \right) \int_0^1 du e^{i \bar u q x} \psi^v(u)
\nonumber \\
&&\langle \gamma(q) \vert \bar q(x) \gamma_\mu \gamma_5 q(0) \vert 0
\rangle  = - \frac{1}{4} e_q f_{3 \gamma} \epsilon_{\mu \nu \alpha
\beta } \varepsilon^\nu q^\alpha x^\beta \int_0^1 du e^{i \bar u q
x} \psi^a(u)
\nonumber \\
&&\langle \gamma(q) \vert  \bar q(x) \sigma_{\mu \nu} q(0) \vert  0
\rangle  = -i e_q \langle \bar q q \rangle (\varepsilon_\mu q_\nu - \varepsilon_\nu
q_\mu) \int_0^1 du e^{i \bar u qx} \left(\chi \varphi_\gamma(u) +
\frac{x^2}{16} \mathbb{A}  (u) \right) \nonumber \\ 
&&-\frac{i}{2(qx)}  e_q \bar qq \left[x_\nu \left(\varepsilon_\mu - q_\mu
\frac{\varepsilon x}{qx}\right) - x_\mu \left(\varepsilon_\nu -
q_\nu \frac{\varepsilon x}{q x}\right) \right] \int_0^1 du e^{i \bar
u q x} h_\gamma(u)
\nonumber \\
&&\langle \gamma(q) | \bar q(x) g_s G_{\mu \nu} (v x) q(0) \vert 0
\rangle = -i e_q \langle \bar q q \rangle \left(\varepsilon_\mu q_\nu - \varepsilon_\nu
q_\mu \right) \int {\cal D}\alpha_i e^{i (\alpha_{\bar q} + v
\alpha_g) q x} {\cal S}(\alpha_i)
\nonumber \\
&&\langle \gamma(q) | \bar q(x) g_s \tilde G_{\mu \nu}(v
x) i \gamma_5  q(0) \vert 0 \rangle = -i e_q \langle \bar q q \rangle \left(\varepsilon_\mu q_\nu -
\varepsilon_\nu q_\mu \right) \int {\cal D}\alpha_i e^{i
(\alpha_{\bar q} + v \alpha_g) q x} \tilde {\cal S}(\alpha_i)
\nonumber \\
&&\langle \gamma(q) \vert \bar q(x) g_s \tilde G_{\mu \nu}(v x)
\gamma_\alpha \gamma_5 q(0) \vert 0 \rangle = e_q f_{3 \gamma}
q_\alpha (\varepsilon_\mu q_\nu - \varepsilon_\nu q_\mu) \int {\cal
D}\alpha_i e^{i (\alpha_{\bar q} + v \alpha_g) q x} {\cal
A}(\alpha_i)
\nonumber \\
&&\langle \gamma(q) \vert \bar q(x) g_s G_{\mu \nu}(v x) i
\gamma_\alpha q(0) \vert 0 \rangle = e_q f_{3 \gamma} q_\alpha
(\varepsilon_\mu q_\nu - \varepsilon_\nu q_\mu) \int {\cal
D}\alpha_i e^{i (\alpha_{\bar q} + v \alpha_g) q x} {\cal
V}(\alpha_i) \nonumber\\
&& \langle \gamma(q) \vert \bar q(x)
\sigma_{\alpha \beta} g_s G_{\mu \nu}(v x) q(0) \vert 0 \rangle  =
e_q \langle \bar q q \rangle \left\{
        \left[\left(\varepsilon_\mu - q_\mu \frac{\varepsilon x}{q x}\right)\left(g_{\alpha \nu} -
        \frac{1}{qx} (q_\alpha x_\nu + q_\nu x_\alpha)\right) \right. \right. q_\beta
\nonumber \\
 && -
         \left(\varepsilon_\mu - q_\mu \frac{\varepsilon x}{q x}\right)\left(g_{\beta \nu} -
        \frac{1}{qx} (q_\beta x_\nu + q_\nu x_\beta)\right) q_\alpha
-
         \left(\varepsilon_\nu - q_\nu \frac{\varepsilon x}{q x}\right)\left(g_{\alpha \mu} -
        \frac{1}{qx} (q_\alpha x_\mu + q_\mu x_\alpha)\right) q_\beta
\nonumber \\
 &&+
         \left. \left(\varepsilon_\nu - q_\nu \frac{\varepsilon x}{q.x}\right)\left( g_{\beta \mu} -
        \frac{1}{qx} (q_\beta x_\mu + q_\mu x_\beta)\right) q_\alpha \right]
   \int {\cal D}\alpha_i e^{i (\alpha_{\bar q} + v \alpha_g) qx} {\cal T}_1(\alpha_i)
\nonumber \\
&&+
        \left[\left(\varepsilon_\alpha - q_\alpha \frac{\varepsilon x}{qx}\right)
        \left(g_{\mu \beta} - \frac{1}{qx}(q_\mu x_\beta + q_\beta x_\mu)\right) \right. q_\nu
\nonumber \\ &&-
         \left(\varepsilon_\alpha - q_\alpha \frac{\varepsilon x}{qx}\right)
        \left(g_{\nu \beta} - \frac{1}{qx}(q_\nu x_\beta + q_\beta x_\nu)\right)  q_\mu
\nonumber \\ 
 && -
         \left(\varepsilon_\beta - q_\beta \frac{\varepsilon x}{qx}\right)
        \left(g_{\mu \alpha} - \frac{1}{qx}(q_\mu x_\alpha + q_\alpha x_\mu)\right) q_\nu
\nonumber 
\end{eqnarray*}
\begin{eqnarray*}
 &&+
         \left. \left(\varepsilon_\beta - q_\beta \frac{\varepsilon x}{qx}\right)
        \left(g_{\nu \alpha} - \frac{1}{qx}(q_\nu x_\alpha + q_\alpha x_\nu) \right) q_\mu
        \right]      
    \int {\cal D} \alpha_i e^{i (\alpha_{\bar q} + v \alpha_g) qx} {\cal T}_2(\alpha_i)
\nonumber \\
&&+\frac{1}{qx} (q_\mu x_\nu - q_\nu x_\mu)
        (\varepsilon_\alpha q_\beta - \varepsilon_\beta q_\alpha)
    \int {\cal D} \alpha_i e^{i (\alpha_{\bar q} + v \alpha_g) qx} {\cal T}_3(\alpha_i)
\nonumber \\ &&+
        \left. \frac{1}{qx} (q_\alpha x_\beta - q_\beta x_\alpha)
        (\varepsilon_\mu q_\nu - \varepsilon_\nu q_\mu)
    \int {\cal D} \alpha_i e^{i (\alpha_{\bar q} + v \alpha_g) qx} {\cal T}_4(\alpha_i)
                        \right\}~,
\end{eqnarray*}
where $\varphi_\gamma(u)$ is the DA of leading twist-2, $\psi^v(u)$,
$\psi^a(u)$, ${\cal A}(\alpha_i)$ and ${\cal V}(\alpha_i)$, are the twist-3 amplitudes, and
$h_\gamma(u)$, $\mathbb{A}(u)$, ${\cal S}(\alpha_i)$, ${\cal{\tilde S}}(\alpha_i)$, ${\cal T}_1(\alpha_i)$, ${\cal T}_2(\alpha_i)$, ${\cal T}_3(\alpha_i)$ 
and ${\cal T}_4(\alpha_i)$ are the
twist-4 photon DAs.
The measure ${\cal D} \alpha_i$ is defined as

\begin{eqnarray*}
\label{nolabel05}
\int {\cal D} \alpha_i = \int_0^1 d \alpha_{\bar q} \int_0^1 d
\alpha_q \int_0^1 d \alpha_g \delta(1-\alpha_{\bar
q}-\alpha_q-\alpha_g)~.\nonumber
\end{eqnarray*}

The expressions of the DAs that are entered into the matrix elements above are  as follows:
\begin{eqnarray}
\varphi_\gamma(u) &=& 6 u \bar u \left( 1 + \varphi_2(\mu)
C_2^{\frac{3}{2}}(u - \bar u) \right),
\nonumber \\
\psi^v(u) &=& 3 \left(3 (2 u - 1)^2 -1 \right)+\frac{3}{64} \left(15
w^V_\gamma - 5 w^A_\gamma\right)
                        \left(3 - 30 (2 u - 1)^2 + 35 (2 u -1)^4
                        \right),
\nonumber \\
\psi^a(u) &=& \left(1- (2 u -1)^2\right)\left(5 (2 u -1)^2 -1\right)
\frac{5}{2}
    \left(1 + \frac{9}{16} w^V_\gamma - \frac{3}{16} w^A_\gamma
    \right),
\nonumber \\
h_\gamma(u) &=& - 10 \left(1 + 2 \kappa^+\right) C_2^{\frac{1}{2}}(u
- \bar u),
\nonumber \\
\mathbb{A}(u) &=& 40 u^2 \bar u^2 \left(3 \kappa - \kappa^+
+1\right)  +
        8 (\zeta_2^+ - 3 \zeta_2) \left[u \bar u (2 + 13 u \bar u) \right.
\nonumber \\ && + \left.
                2 u^3 (10 -15 u + 6 u^2) \ln(u) + 2 \bar u^3 (10 - 15 \bar u + 6 \bar u^2)
        \ln(\bar u) \right],
\nonumber \\
{\cal A}(\alpha_i) &=& 360 \alpha_q \alpha_{\bar q} \alpha_g^2
        \left(1 + w^A_\gamma \frac{1}{2} (7 \alpha_g - 3)\right),
\nonumber \\
{\cal V}(\alpha_i) &=& 540 w^V_\gamma (\alpha_q - \alpha_{\bar q})
\alpha_q \alpha_{\bar q}
                \alpha_g^2,
\nonumber \\
{\cal T}_1(\alpha_i) &=& -120 (3 \zeta_2 + \zeta_2^+)(\alpha_{\bar
q} - \alpha_q)
        \alpha_{\bar q} \alpha_q \alpha_g,
\nonumber \\
{\cal T}_2(\alpha_i) &=& 30 \alpha_g^2 (\alpha_{\bar q} - \alpha_q)
    \left((\kappa - \kappa^+) + (\zeta_1 - \zeta_1^+)(1 - 2\alpha_g) +
    \zeta_2 (3 - 4 \alpha_g)\right),
\nonumber \\
{\cal T}_3(\alpha_i) &=& - 120 (3 \zeta_2 - \zeta_2^+)(\alpha_{\bar
q} -\alpha_q)
        \alpha_{\bar q} \alpha_q \alpha_g,
\nonumber \\
{\cal T}_4(\alpha_i) &=& 30 \alpha_g^2 (\alpha_{\bar q} - \alpha_q)
    \left((\kappa + \kappa^+) + (\zeta_1 + \zeta_1^+)(1 - 2\alpha_g) +
    \zeta_2 (3 - 4 \alpha_g)\right),\nonumber \\
{\cal S}(\alpha_i) &=& 30\alpha_g^2\{(\kappa +
\kappa^+)(1-\alpha_g)+(\zeta_1 + \zeta_1^+)(1 - \alpha_g)(1 -
2\alpha_g)\nonumber +\zeta_2[3 (\alpha_{\bar q} - \alpha_q)^2-\alpha_g(1 - \alpha_g)]\},\nonumber \\
\tilde {\cal S}(\alpha_i) &=&-30\alpha_g^2\{(\kappa -\kappa^+)(1-\alpha_g)+(\zeta_1 - \zeta_1^+)(1 - \alpha_g)(1 -
2\alpha_g)\nonumber +\zeta_2 [3 (\alpha_{\bar q} -\alpha_q)^2-\alpha_g(1 - \alpha_g)]\}.
\end{eqnarray}

The numerical values of the parameters used in the DAs are $\varphi_2(1~GeV) = 0$, 
$w^V_\gamma = 3.8 \pm 1.8$, $w^A_\gamma = -2.1 \pm 1.0$, $\kappa = 0.2$, $\kappa^+ = 0$, $\zeta_1 = 0.4$, and $\zeta_2 = 0.3$.

 \end{widetext}
  
  \bibliographystyle{elsarticle-num}
\bibliography{Bottom-charm.bib}
\end{document}